\documentclass{amsart}
\usepackage{amssymb,amscd}
\usepackage{amsthm}

\def\torus{{\bf T}}
\def\integer{{\bf Z}}
\def\reals{{\bf R}}
\def\complex{{\bf C}}

\def\alg{{\mathcal A}}
\def\bdd{{\mathcal B}}
\def\cpt{{\mathcal K}}
\def\cS{{\mathcal S}}
\def\cL{{\mathcal L}}

\def\wt#1{\widetilde{#1}}
\def\ft#1{\widehat{#1}}
\def\conj#1{\overline{#1}}

\newcommand\Aut{\operatorname{Aut}}
\newcommand\Ca{$C^*$-algebra}

\newcommand{\thm}[1]{\advance\count32 by 1\bigskip\noindent{\bf #1 \the\count31.\the\count32.}}

\theoremstyle{plain}
\newtheorem{theorem}{Theorem}[section]

\newtheorem{proposition}[theorem]{Proposition}
\newtheorem{corollary}[theorem]{Corollary}

\theoremstyle{definition}
\newtheorem{definition}[theorem]{Definition}
\newtheorem{definition*}{Definition}
\newtheorem{example}[theorem]{Example}
\theoremstyle{remark}

\newtheorem{remarks*}{Remarks}

\begin{document}

\title[NCP torus bundles via parametrised deformation quantization]
{Noncommutative principal torus bundles via \\ parametrised
strict deformation quantization}

\author[KC Hannabuss]{Keith C. Hannabuss}

\address[Keith Hannabuss]{
Mathematical Institute, 24-29 St. Giles', Oxford, OX1 3LB, and
Balliol College, Oxford, OX1 3BJ,
England}
\email{kch@balliol.oxford.ac.uk}

\author[V Mathai]{Varghese Mathai}

\address[Varghese Mathai]{
Department of Pure Mathematics,
University of Adelaide,
Adelaide, SA 5005,
Australia}
\email{mathai.varghese@adelaide.edu.au}

\begin{abstract}
In this paper, we initiate the study of a parametrised version of Rieffel's strict deformation quantization.
We apply it to give a classification of noncommutative principal torus bundles,
in terms of parametrised strict deformation quantization of ordinary principal torus bundles.
The paper also contains a putative definition of noncommutative non-principal torus bundles.
\end{abstract}

\thanks{{\em Acknowledgements}. V.M. thanks the Australian Research Council for support.}

\maketitle

\section*{Introduction}

Operator theoretic deformation quantization appeared in quantum physics 
a long time ago, but was put on a firm footing relatively recently 
by Rieffel \cite{Rieffel1} (see the references therein), who called it 
{\em strict} deformation quantization, mainly to distinguish it from {\em formal} deformation
quantization, where convergence isn't an issue. His theory has been remarkably 
successful, giving rise to many  examples of noncommutative manifolds, 
which have become extremely useful both in mathematics and mathematical physics.
In a recent paper \cite{ENOO}  Echterhoff, Nest, and Oyono-Oyono defined
noncommutative principal torus bundles, inspired by fundamental results in \cite{Rieffel3},
as well as the T-duals of certain continuous trace algebras \cite{MR, MR2}.
They also classified all noncommutative principal torus bundles in terms of (noncommutative)
fibre products of principal torus bundles and group C$^*$-algebras of lattices in simply-connected 2-step nilpotent Lie groups, cf. \S\ref{sect:classify}. In this paper, we show that their classification
can be neatly understood in terms of a generalization of Rieffel's strict deformation quantization \cite{Rieffel1,Rieffel2},
to the parametrised case that is developed here.
More precisely, we generalize to the parametrised case, the recent version of Rieffel's strict deformation quantization given by Kasprzak \cite{Kasp} based on work of Landstad \cite{Land1, Land2}. More precisely, we give a classification of noncommutative principal torus bundles,
in terms of parametrised strict deformation quantization of ordinary principal torus bundles.

Strict deformation quantization theory works with smooth
subalgebras, so we start with a section, \S\ref{sec:prelim}, on  smooth subalgebras
of C$^*$-algebras and a smooth version of the noncommutative
torus bundle theory. That is followed in \S\ref{sect:rieffel} by a summary  recalling
the ideas of Rieffel's strict deformation quantization, and
then a section, \S\ref{sect:parameter}, generalising that to a parametrised version.
Then we explain in \S\ref{sect: Landstad},  Kasprzak's recent account of 
Rieffel's strict deformation quantization theory
based on ideas of Landstad, and extend it to a  parametrised
version.  In \S\ref{sect:classify}, after a summary of the relevant parts of the
Echterhoff, Nest, and Oyono-Oyono classification of
noncommutative principal torus bundles, we explain the
connection to parametrised strict deformation quantization
theory. We end with a section, \S\ref{sect:examples}, containing examples of parametrised 
strict deformation quantization including the case of principal torus bundles.
It also contains a putative definition of noncommutative 
non-principal torus bundles. There is an appendix containing a discussion about factors of automorphy which
is used in the paper.

\section{Fibrewise smooth $*$-bundles}
\label{sec:prelim}
We begin by recalling the notion of $C^*$-bundles over $X$ and the special
case of noncommutative principal bundles. Then we discuss the fibrewise smoothing
of these, which is used in parametrised Rieffel deformation later on.

Let $X$ be a locally compact Hausdorff space and let $C_0(X)$ denote
the  {\Ca} of continuous functions on $X$ that vanish at infinity.
A $C^*$-\emph{bundle} $A(X)$ over $X$ in the sense of \cite{ENOO}
is  exactly a $C_0(X)$-algebra
in the sense of Kasparov \cite{Kas88}.  That is, $A(X)$ is a
{\Ca} together with a non-degenerate $*$-homomorphism
$$\Phi_A: C_0(X)\to ZM(A(X)),$$ called the {\em structure map}, where
$ZM(A)$ denotes the center of the multiplier algebra $M(A)$ of $A$.
The {\em fibre}   over  $x\in X$ is then  $A(X)_x=A(X)/I_x$, where
$$I_x=\{\Phi(f)\cdot a;\, a\in A(X) \text{ and } f\in C_0(X)\text{
such that } f(x)=0\},$$ and the canonical quotient map $q_x:A(X)\to
A(X)_x$ is called the {\em evaluation map}  at $x$.

Note that this definition does not require local triviality of the
bundle, or even for the fibres of the bundle to be isomorphic to one
another.

Let $G$ be a locally compact group. One says that there is a {\em
fibrewise action} of $G$ on a $C^*$-bundle $A(X)$ if there is a
homomorphism $\alpha: G \longrightarrow  \Aut(A(X))$ which is
$C_0(X)$-linear in the sense that
$$
\alpha_g(\Phi(f) a) = \Phi(f)(\alpha_g(a)), \qquad \forall g\in G, \,
a\in A(X),\, f \in C_0(X).
$$
This means that $\alpha$ induces an action $\alpha^x$ on the fibre
$A(X)_x$ for all $x\in X$.

The first observation is that if
$A(X)$ is a $C^*$-algebra bundle over $X$ with a fibrewise action $\alpha$
of a {\em Lie group} $G$, then there is a canonical {\em smooth $*$-algebra bundle}
over $X$. We recall its definition from \cite{Co80}. 
A vector $y\in A(X)$ is said to be a {\em smooth vector} if the map
$$
G \ni g \longrightarrow \alpha_g(y) \in A(X)
$$
is a smooth map from $G$ to the normed vector space $A(X)$. Then
$$
\alg^\infty(X) = \{ y\in A(X)\, | \, y\, \text{ is a smooth vector} \}
$$
is a $*$-subalgebra of $A(X)$ which is norm dense in $A(X)$. Since $G$ 
acts {\em fibrewise} on $A(X)$, it follows that $\alg^\infty(X) $ is again a
$C_0(X)$-algebra which is {\em fibrewise smooth}.

Let $T$ denote the torus of dimension $n$.  The authors of
\cite{ENOO} define a  {\em noncommutative principal  $T$-bundle}
(or {\em NCP $T$-bundle}) over $X$ to be a separable  $C^*$-bundle
$A(X)$  together with a fibrewise  action $\alpha:T\to\Aut(A(X))$
such that  there is a Morita equivalence,
$$A(X)\rtimes_{\alpha}T\cong C_0(X, \cpt),$$ 
as $C^*$-bundles over $X$,
where $ \cpt$ denotes the $C^*$-algebra of compact operators.

The motivation for calling such $C^*$-bundles $A(X)$ NCP
$T$-bundles arises from a special case of a theorem of
Rieffel \cite{Rieffel3}, which
states that if $q: Y \longrightarrow X$  is a principal
$T$-bundle, then $C_0(Y) \rtimes T$ is Morita equivalent to $C_0(X,\cpt)$.

If $A(X)$ is a NCP $T$-bundle over $X$, then we call $\alg^\infty(X)$
a {\em fibrewise smooth noncommutative principal  $T$-bundle}
(or {\em fibrewise smooth NCP $T$-bundle}) over $X$. In this paper, we are able
to give a complete classification of fibrewise smooth NCP $T$-bundles over $X$
via a parametrised version of Rieffel's theory of strict deformation quantization.

\section{Rieffel deformation}
\label{sect:rieffel}

Unlike Rieffel's deformation theory \cite{Rieffel1,Rieffel2},
the version which we shall use \cite{Land1,Land2,Kasp} starts
with multipliers, so in this section we shall recapitulate some
standard results but in a formulation which suits the later
extension to a parametrised theory and the Landstad--Kasprzak
approach.  In what follows, a Poisson bracket $\{,\}$ on $\alg$
is a bilinear form from $\alg$ to itself, which is a Hochschild
2-cocycle satisfying a couple of additional technical
conditions that will not be repeated here, but we refer the
reader to  \S5 in \cite{Rieffel1}.

\begin{definition}[\S5, \cite{Rieffel1}]
Let $\alg$ be a dense $*$-subalgebra of a C$^*$ algebra, equipped with a Poisson bracket $\{,\}$.
A {\it strict deformation quantisation of $\alg$ in the direction of $\{,\}$} means an open interval $I$ containing $0$ in $\reals$, together with associative products $\star_\hbar, \, \hbar\in I$, involutions and C$^*$-norms on $\alg$ which for $\hbar = 0$ are the original product, involution and norm on $\alg$, such that:
\begin{enumerate}
\item
The   corresponding field of C$^*$-algebras with continuity structure given by the elements of $\alg$ as constant fields, is a continuous field of C$^*$-algebras.
\item
For all $a,b \in \alg$, as $\hbar \to 0$ one has $\|(a\star_{\hbar} b - ab)/(i\hbar) - \{a,b\}\| \to 0$.
\end{enumerate}
\end{definition}

Typically, one tries to find strict deformation quantizations of Poisson manifolds,
thus obtaining interesting noncommutative manifolds.

Rieffel's definition and construction are motivated by Moyal's
product but to link it with Kasprzak's work it is useful to
give the background.

Suppose that $\alg$ is a pre- C$^*$-algebra
with an action $\alpha$ of a locally compact abelian group $V$ (written
additively), and let $\sigma$ be a multiplier on its Pontryagin dual $\ft{V}$,
that $\sigma: \ft{V}\times \ft{V} \to \torus$ is a borel map, satisfying the
cocycle identity
$$
\sigma(\xi,\eta)\sigma(\xi+\eta,\zeta) =
\sigma(\xi,\eta+\zeta)\sigma(\eta,\zeta),
$$
for all $\xi, \eta, \zeta \in \ft{V}$. The group of all such cocycles (or multipliers) is denoted by $Z^2(\ft{V},\torus)$. Two multipliers $\sigma_1$ and $\sigma_2$ are equivalent (or cohomologous)  if and only if there is a borel map $\rho: \ft{V} \to
\torus$ such that
$$
\sigma_1(\xi,\eta) \rho(\xi+\eta) = \sigma_2(\xi,\eta) \rho(\xi)\rho(\eta),
$$
and the equivalence classes form the cohomology group $H^2(\ft{V},\torus)$.  A cocycle equivalent to the constant cocycle $\ft{V}\times \ft{V} \to \{1\}$ is said to be trivial.

Recalling that a bicharacter $\beta: V\times V \to \torus$ defines characters $\beta^1_\xi: \eta \mapsto \beta(\xi,\eta)$ for each fixed $\xi$, and $\beta^2_\eta: \xi\mapsto \beta(\xi,\eta)$ for fixed $\eta$, we see that bicharacters always define cocycles, because
$$
\beta(\xi,\eta)\beta(\xi+\eta,\zeta) =
\beta(\xi,\eta)\beta(\xi,\zeta)\beta(\eta,\zeta) =
\beta(\xi,\eta+\zeta)\beta(\eta,\zeta).
$$

\begin{theorem}[\cite{Kl,KCH}]
 Every multiplier on an abelian group $\ft{V}$ is equivalent to a bicharacter (and so is continuous).
Two bicharacters $\beta_1$ and $\beta_2$ are equivalent  if and only if
$\beta = \beta_1\beta_2^{-1}$ a symmetric bicharacter , that is $\beta(\xi,\eta) = \beta(\eta,\xi)$.
If $\ft{V} = 2\ft{V}$ then each cohomology class can be represented by a unique antisymmetric bicharacter $\beta$, that is $\beta(\xi,\eta)= \beta(\eta,\xi)^{-1}$.
\end{theorem}

We can therefore assume that $\sigma$ is a bicharacter, and
this means that it is actually continuous in each variable.
When $\ft{V} = 2\ft{V}$,  the element
$\sigma(\xi,\frac12\eta)/\sigma(\eta,\frac12\xi)$ gives the
canonical antisymmetric bicharacter representative of the class
containing $\sigma$. We note that vector groups $\ft{V} =
\reals^n = 2\ft{V}$, so that each cocycle can be represented by
a continuous antisymmetric antisymmetric bicharacter, which
must be the exponential $\exp[i\pi  s(\xi,\eta)]$ of a
skew-symmetric bilinear form $s$. These can be identified with
$\bigwedge^2 V$. There is a similar analysis for lattices $L
\cong \integer^n$, where the bicharacters are given by the
torus $\bigwedge^2(V/L) = \bigwedge^2V/\bigwedge^2L$
(restrictions modulo those with trivial restriction), but a
torus $\torus^n$ has only trivial bicharacters $\beta(\xi,\eta)
= 1$, due to the following observation.

\begin{corollary}
There are no non-trivial bicharacters on a connected compact group $V$.
\end{corollary}

This follows because the non-trivial multipliers on the dual of an infinite connected 
compact
group (such as $V=\torus^{2n}$) are never invertible, since the
dual (e.g. $\ft{V} = \integer^{2n}$) is discrete and the two
groups are not isomorphic.

\begin{theorem}
Given a  continuous bicharacter cocycle $\sigma$  on $\ft{V}$
and a pre-$C^*$ algebra $\alg$ we may form the $*$-algebra of
functions $f, g: \ft{V} \to \alg$ smooth with respect to the
translation automorphisms $\tau_u[f] = f(u+v)$,  with the
twisted convolution product
$$
(f * g)(\xi) = \int_{\ft{V}} \sigma(\eta,\xi-\eta)
f(\eta)g(\xi-\eta)\,d\eta,
$$
and involution $f^*(\xi) = \sigma(\xi,\xi)\conj{f(-\xi)}$.
Up to isomorphism this algebra depends only on the cohomology class of $\sigma$.
\end{theorem}

\begin{proof}
The cocycle identity on $\sigma$ ensures associativity. When
$\sigma$ is an antisymmetric bicharacter the involution reduces
to  $f^*(\xi) =  \conj{f(-\xi)}$.  Changing $\sigma$ to 
$$
\sigma(\xi,\eta) \rho(\xi)\rho(\eta) \rho(\xi+\eta)^{-1}$$ gives
the algebra isomorphism $f \mapsto \rho.f$ (the pointwise
product).
\end{proof}

\bigskip
These functions can be Fourier transformed to functions on $V$
$$
\ft{f}(v) = \int_{\ft{V}} \xi(v) f(\xi)\,d\xi,
$$
where we assume that the Haar measure is normalised to make the transform
unitary, and one has the usual inverse transform.

\begin{theorem}
The transformed product is
$$
\ft{(f * g)}(v) = \int \sigma(\eta,\xi)\conj{\eta(u)}\conj{\xi(w)}\ft{f}(v+u)\ft{g}(v+w)\,dudwd\xi
d\eta.
$$
\end{theorem}
\begin{proof}
We calculate that
\begin{align*}
\ft{(f * g)}(v) &= \int_{\ft{V}} \xi(v)(f\star g)(\xi)\,d\xi \\
&=\int_{\ft{V}\times\ft{V}}
\xi(v)\sigma(\eta,\xi\eta^{-1})f(\eta)g(\xi\eta^{-1})\,d\xi d\eta\\
&=
\int_{\ft{V}\times\ft{V}} \sigma(\eta,\xi) \eta(v)f(\eta)\xi(v)g(\xi)\,d\xi
d\eta \\
&= \int
\sigma(\eta,\xi)\eta(v)\ft{f}(u)\conj{\eta(u)}\xi(v)\ft{g}(w)\conj{\xi(w)}\,dudwd\xi
d\eta \\
 &= \int \sigma(\eta,\xi)\conj{\eta(u)}\conj{\xi(w)}\ft{f}(v+u)\ft{g}(v+w)\,dudwd\xi
d\eta,\\
\end{align*}
where we replaced $u$ and $w$ by $v+u$ and $v+w$ in the last step.
\end{proof}

We now want to connect this transformed product with Rieffel's
deformation. To this end  we introduce a bicharacter $e$ on
$V$, which defines a homomorphism $e^1: V \to \ft{V}$. Rieffel
works with a vector group $V$ and $e(u,w) = \exp(i(u\cdot w))$
for some inner product on $V$. When $\sigma$ is non-degenerate
(that is, $\sigma^1:\ft{V} \to V$ is an isomorphism) we can
choose $e$ so that $e^1$ is the inverse of $\sigma^1$, but in
general we have an automorphism $T = \sigma^1\circ e^1: V \to
V$. As a final piece of notation we introduce the adjoint $T^*$
with respect to $e$: $e(T^*u,w) = e(u,Tw)$.

\begin{proposition}\label{prop:bichar}
The bicharacters $\sigma$ and $e$ are related by
$\sigma(e^1_u,e^1_v) = e^1_v(Tu) = e(Tu,v)$ for all $u,v\in V$.
Suppose that $\sigma$ is an antisymmetric bicharacter. Then if
$e$ is symmetric $T=-T^*$, and if $e$ is antisymmetric $T =
T^*$.
\end{proposition}

\begin{proof}
By definition we have
$$
\sigma(e^1_u,e^1_v) = e^1_v(Tu) = e(Tu,v).
$$
Since $\sigma$ is skew symmetric this gives
$$
e(Tu,v) = \sigma(e^1_u,e^1_v) = \sigma(e^1_v,e^1_u)^{-1} = e(Tv,u)^{-1} = e(-Tv,u).
$$
When $e$ is symmetric this shows that $e(Tu,v) = e(u,-Tv)$, so that $T = -T^*$, and when $e$ is antisymmetric $T= T^*$.
\end{proof}

\begin{theorem}
Given a non-degenerate bicharacter $e$ on $V$, set $T = \sigma^1\circ e^1: V \to V$.
and  $e(T^*u,w) = e(u,Tw)$.
Then
$$
\ft{(f * g)}(v) =  \int e(u,w)\ft{f}(v+T^*u)\ft{g}(v+w)\,dudw.
$$
\end{theorem}

We change the order of integration in our earlier expression for $f*g$ and concentrate on the integrals over $\ft{V}$:
$$
\int_{\ft{V}\times\ft{V}} \sigma(\eta,\xi)\conj{\eta(u)}\conj{\xi(w)}\,d\xi
d\eta.
$$
Since $e$ is nondegenerate we may set $\eta = e_v$, and then, by definition, we
have
$$
\sigma(e_v,\xi)\conj{e_v(u)}\conj{\xi(w)} = \xi(Tv)\conj{e(v,u)}\conj{\xi(w)}
$$
By the Fourier inversion theorem, integration over $\xi$ gives a delta function
$\delta(Tv - w)$. Replacing $u$ by $T^*u$ and integrating over $v$, we now get
$$
\int \sigma(e_v,\xi)\conj{e(v,T^*u)}\conj{\xi(w)}\,d\xi dv = \int
\delta(w-Tv)e(Tv,u)\,dv = e(w,u),
$$
Up to a multiple, integration over $\eta$ and $v$ are the same,
and with appropriate choices of measure we can ensure that they
agree precisely. Then inserting this into the original formula
for the product we have
$$
\ft{(f * g)}(v) =  \int e(u,w)\ft{f}(v+T^*u)\ft{g}(v+w)\,dudw.
$$

\begin{theorem}
The Fourier transformed product $\ft{(f * g)} = \ft{f}\star\ft{g}$ where
$$
(\ft{f}\star\ft{g})(v) =  \int e(u,w)\ft{f}(v+T^*u)\ft{g}(v+w)\,dudw.
$$
In terms of the translation automorphisms $\tau_w[g](v) = g(v+w)$, we have
$$
(\ft{f}\star\ft{g})(v) =  \int
e(u,w)\tau_{T^*u}[\ft{f}](v)\tau_w[\ft{g}](v)\,dudw.
$$
\end{theorem}

Evaluating at the identity $v=0$ gives
$$
(\ft{f}\star\ft{g})(0) =  \int
e(u,w)\tau_{T^*u}[\ft{f}](0)\tau_w[\ft{g}](0)\,dudw.
$$
Rieffel noticed that this formula can now be interpreted whenever $\alpha$
defines automorphisms of $\alg$, so that one can define
$$
a\star b =  \int e(u,w)\alpha_{T^*u}[a]\alpha_w[b]\,dudw,
$$
for $a$ and $b$ in the algebra. (Our $T^*$ is Rieffel's $J$.)
When both bicharacters $\sigma$ and $e$ are nondegenerate we can also write this as
$$
a\star b =  \int \det[T^*]^{-1}e({T^*}^{-1}u,w)\alpha_{u}[a]\alpha_w[b]\,dudw.
$$

The above arguments are formal and one must check that the integrals converge.
In the standard Moyal theory this is done by working only with Schwarz
functions and in the general case one uses the smooth vectors $\alg^\infty$ for
the action $\alpha$, which form a dense Fr\'echet subalgebra of $\alg$.

For vector groups this works particularly smoothly, and one
obtains a strict deformation quantisation \cite{Rieffel1},
Theorem 9.3. However,  there are technical problems when  $V =
\torus^{2n}$ since, as we have seen,  there are no nontrivial
bicharacters $e$ on $V$. There are two ways of dealing with
this problem. One is by the Kasprzak--Landstad approach of
working with the dual crossed product algebra,
\cite{Land2,Kasp}, and the other is Rieffel's approach of
lifting the action of the torus $T = V/L$ (with $L$ a lattice),
to the vector group $V$, \cite{Rieffel1} Ch 2.

\section{Parametrised Rieffel deformations}
\label{sect:parameter}

An interesting generalisation comes from inserting a parameter.
More precisely, we work with a $C_0(X)$-algebra $\alg$, where
$X$ is a locally compact Hausdorff space. (That is there is a
map $C_0(X) \to {\mathcal ZM}\alg$.) Consider a function $\sigma \in
C_b(X, Z^2(\ft{V},\torus))$ taking values in the bicharacter
cocycles . At each point $x \in X$ this defines a multiplier
$\sigma_x$, and a map $\sigma^1_x: \ft{V} \to V$. We then form $T_x =
\sigma^1_x\circ e^1$ and its adoint $T^*_x$ with respect to $e$, where 
$e, e^1$ are defined just prior to Proposition \ref{prop:bichar}. If
the image of $\sigma$ lies in the non-degenerate cocycles we can
then form the continuous function $x \mapsto
e({T_x^*}^{-1}u,v)$, which acts on $\alg$.

\begin{theorem}
\label{thm:parameter}
Given a $C_0(X)$-algebra $\alg$, where $X$ is a locally compact
Hausdorff space, and a function $\sigma \in C_b(X,
Z^2(\ft{V},\torus))$ taking values in the nondegenerate
bicharacter cocycles , let $T_x = \sigma^1_x\circ e^1$ and
$e(T_x^*u,w) = e(u,T_xw)$.  Then, if $T$ has an inverse in a
subalgebra of $C_b(X)$ whose action preserves the (fibrewise) smooth
subalgebra $\alg^\infty$, one has a product
$$
a\star b =  \int \det[T^*]^{-1}e({T^*}^{-1}u,w)\alpha_{u}[a]\alpha_w[b]\,dudw,
$$
defined by the actions of the continuous functions
$\det[T^*]^{-1}$ and $e({T^*}^{-1}u,w)$ on $\alg$. This gives
an algebra $\alg_\sigma$ with an involution, which  inherits a
$C_0(X)$-algebra structure. (The $C_0(X)$-structure on the
algebra is such that for $F \in C_0(X)$ we have
$F.(a\star b) = (F.a)\star b = a\star(F.b)$.)
\end{theorem}

For vector groups it follows from the definition that the iterated parametrised strict deformation quantization
$(\alg_{\sigma_1})_{\sigma_2} \cong \alg_{\sigma_1\sigma_2}$, with the isomorphism
defined by the obvious identification map. This follows on
writing down the repeated deformation product and evaluating a
double integral using Parseval's formula or the Fourier
inversion formula. Alternatively we can note that the
bicharacter $\sigma^{-1}$ can always be written in Rieffel form, and
then the result follows from his. Yet another approach would be
to use the equivalence with Kasprzak's formulation given below,
and then to deduce it from his result. In particular, we can
undeform $\alg_\sigma$ using $\conj{\sigma}$.

In this more general context we can generalise Rieffel's
discussion  of the action of continuous automorphisms of the
group $V$ (which give $GL(V)$ when $V$ is a vector group), to
allow functions $S \in C^\infty(X,Aut(V))$ and using
$$
a\star_S b = \int_{V\times V} \sigma^{-1}(Su,Sw).(\alpha_{u}[a] \alpha_{w}[b])\,dudw.
$$

Note that the original automorphisms of $V$ on $\alg$ are also
automorphisms of the deformed algebra, since
\begin{eqnarray*}
\alpha_v[a]\star_\sigma\alpha_v[b] &=&
\int_{V\times V}
\det[T^*]^{-1}e({T^*}^{-1}u,v).(\alpha_{u+v}[a]
\alpha_{w+v}[b])\,dudw\\ &=& \int_{V\times V}
\det[T^*]^{-1}e({T^*}^{-1}u,v).\alpha_v[(\alpha_{u}[a]
\alpha_{w}[b])\,dudw\cr &=& \alpha_v[a\star_\sigma b],
\end{eqnarray*}
since $\alpha_w$ commutes with the $C_0(X)$ action.

We constructed the deformation as the dual of a twisted crossed
product, and the reverse is also true. Given an algebra $\alg$
with an action of $V$ one can take the crossed product
$\alg\rtimes V$ with a dual action of $\ft{V}$. Looking first
at the unparametrised case, when $\sigma$ is non-degenerate
there is a dual multiplier $\ft{\sigma}$ on $V$ defined by
$\ft{\sigma}(u,\sigma_1\eta) = \eta(u)$, and similarly for
$\ft{e}(\xi,e^1v) = \xi(v)$, and $\ft{T} =
\ft{\sigma}^1\circ\ft{e}^1$. These definitions effectively mean
that $\ft{e}^1$ is the inverse of $e^1$ and similarly for
$\sigma$. We can now deform the crossed product.
\begin{eqnarray*}
(a\star_{\ft{\sigma}}b)(v) &=&  \int
\ft{e}(\xi,\eta)\ft{\alpha}_{\ft{T}\xi}[a]\ft{\alpha}_\eta[b]\,d\xi
d\eta\\ &=&  \int
\ft{e}(\xi,\eta)\ft{\alpha}_{\ft{T}\xi}[a](u)\alpha_u[\ft{\alpha}_\eta[b](v-u)]\,d\xi
d\eta du\\ &=&  \int
\ft{e}(\xi,\eta)(\ft{T}\xi)(u)\eta(v-u)a(u)\alpha_u[b(v-u)]\,d\xi
d\eta du\\ &=&  \int
\eta(\ft{e}^1\xi)\ft{T}\xi)(u)\eta(v-u)a(u)\alpha_u[b(v-u)]\,d\xi
d\eta du.
\end{eqnarray*}
 The integral over $\eta$ gives a delta
function concentrated on $\ft{e}^1\xi = u-v$, or equivalently
where $\xi = e^1(u-v)$, so that the $\xi$ integral then gives
$$
(a\star_{\ft{\sigma}}b)(v)
=  \int  (\ft{T}e^1(u-v))(u)a(u)\alpha_u[b(v-u)]\,, du.
$$
By definition, we have
$$
 \ft{T}e^1 = \ft{\sigma}^1\ft{e}^1e^1 = \ft{\sigma}^1,
 $$
 which leads to the reduction
$$
(a\star_{\ft{\sigma}}b)(v)
=  \int  (\ft{\sigma}^1(u-v))(u)a(u)\alpha_u[b(v-u)]\,, du
=  \int  \ft{\sigma}(u-v,u)a(u)\alpha_u[b(v-u)]\,, du.
$$
This is a twisted crossed product with multiplier. There is a
similar parametrised version.

\section{Landstad--Kasprzak and Rieffel deformation}
\label{sect: Landstad}

Building on work of Landstad \cite{Land1,Land2}, Kasprzak
\cite{Kasp} gives an alternative dual picture of deformation
theory. It is useful to give the equivalence with Rieffel
deformation explicitly, as Kasprzak omits the details. (The
correspondence is not obvious since the algebra elements in
Rieffel's deformation are the same and only the product
changes, whereas in Kasprzak's formulation the deformed and
undeformed algebras are distinct fixed point subalgebras of the
multiplier algebra of the crossed product, with different
actions of $V$. Smoothness or some equivalent is also needed;
Landstad suggests in \cite{Land2} that it is sufficient to use
the Fourier algebra instead of smooth subalgebras.) In the
following account we use Rieffel's notation of $\alpha$ rather
than $\rho$ for the automorphisms.

Landstad showed in \cite{Land1} that when a group $V$ acts on
an algebra $\alg$, the crossed product $\bdd = \alg\rtimes V$
has a coaction which is defined by a homomorphism $\lambda: V
\to {\mathcal UMB}$ (the unitary multiplier algebra).  By
integration $\lambda$ extends to $C(V) \to {\mathcal MB}$.
When $V$ is abelian there is also the dual Takai--Takesaki
action $\ft{\alpha}$ of $\ft{V}$, and these interact by
$\ft{\alpha}_\xi[\lambda_v] = \xi(v)\lambda_v$. By
Takai--Takesaki duality $\bdd\rtimes_{\ft{\alpha}} \ft{V}$ is
isomorphic to $\alg\otimes \cpt$, reconstructing $\alg$ up to
stable equivalence. When $\bdd$ has the Landstad $\lambda$ as
well we can deduce a stronger duality that there is an algebra
$\alg$ with $V$-action $\alpha$ such that $\bdd =
\alg\rtimes_\alpha V$. Kasprzak's idea is that
$\ft{\alpha}_\xi$ can be deformed  by a cocycle $\sigma$ for
$\ft{V}$ to a new action $\ft{\alpha}^\sigma_\xi$.

\begin{theorem}\cite{Kasp}.
Let $(\bdd,\lambda,\ft{\alpha})$ be as above, and $\sigma$ a continuous cocycle for $\ft{V}$.
 Setting  $U_\xi = \lambda(\sigma^1_\xi)$ there is an action of $\ft{V}$ on $\alg$ given by
$$
\ft{\alpha}^\sigma_\xi :b \mapsto U_xi^*\ft{\alpha}_\xi [b]U_\xi,
$$
which also satisfies $\alpha^\sigma_\xi[\lambda_v] = \xi(v)\lambda_v$.
\end{theorem}

\begin{corollary}
There is an algebra $\alg^\sigma$ and $V$-action $\alpha^\sigma$ such that
the crossed product
$\alg^\sigma\rtimes_{\alpha^\sigma} V \cong \bdd =  \alg\rtimes_\alpha V$.
\end{corollary}

The deformed and undeformed algebras can be identified 
with the subalgebras of ${\mathcal MB}$ fixed by the action $\ft{\alpha}$ of $\ft{V}$.

In particular, the undeformed algebra is fixed under the dual
group action on the crossed product given by
$\ft{\alpha}_\xi[a](v) = \xi(v)a(v)$. The fixed points of this
action are distributions concentrated on the group identity
$v = 0$, which make sense as elements of the multiplier
algebra. They give an algebra isomorphic to $\alg$, and this is
just Rieffel's construction as defined above.

For the algebra deformed by $\sigma,$  $U_\xi =
\delta_{\sigma^1_\xi}$, and by the covariance property of
crossed products the adjoint action of $U_\xi$ is the same as
the action of $\alpha_{\sigma^1_\xi}$. We therefore have
$$
\ft{\alpha}^{\sigma}_\xi[a](v)
= \alpha_{\sigma^1_\xi}^{-1}[\ft{\alpha}_\xi)[a(v)]]
= \xi(x)\alpha_{\sigma^1_\xi}^{-1}[a(v)]
$$

Changing variable, the fixed subalgebra,  where
$\ft{\alpha}^{\sigma}_\xi[a]= a$,  therefore consists of
elements $a$ satisfying
$$
\alpha_{\sigma^1_\xi}[a(v)] = \xi(v)a(v).
$$
In the notation of previous sections we set $\xi = e^1_u$ so
that $\sigma^1_\xi = Tu$, and then the condition becomes
$$
\alpha_{Tu}[a(v)] = e(u, v)a(v).
$$
Thus the value of $a(v)$ always lies in a particular eigenspace
of the action $\alpha$. (In particular, when $e$ is an
antisymmetric bicharacter $a(0)$ must be in the fixed point
algebra of $\alpha_T$.) In other words we can characterise the
fixed point algebra elements as the elements whose value at $v$
lies in the relevant spectral subspace $\ker[\alpha_{Tu} -
e(u,v)]$ of the action of $\alpha$.

To get all the eigenspaces we must do a direct integral, or,
for suitably well-behaved functions (the smooth subalgebra), we
set $I(a) = \int a(v) \,dv$.

\begin{theorem}
When $T$ is invertible, the product of fixed point algebra
elements $a$ and $b$ satisfies
$$
I(a*b) = I(a)\star I(b).
$$
\end{theorem}
\begin{proof}
When $T$ is invertible, the product of fixed point algebra
elements is given by
\begin{align*}
I(a*b) &= \int a(u)\alpha_u[b(v-u)] \,dudv\cr
&= \int a(u)\alpha_u[b(v)] \,dudv\cr
&= \int e(T^{-1}u,v)a(u)b(v) \,dudv.\cr
\end{align*}

On the other hand, under the same conditions, and with $e$ symmetric, so that $T = -T^*$
\begin{align*}
(I(a)\star I(b)) &= \int \det[T^*]^{-1}e({T^*}^{-1}u,v)\alpha_u[ I(a)]\alpha_v[I(b)]\,dudv\cr
&= \int \det[T^*]^{-1}e(-T^{-1}u,v)\alpha_u[ a(y)]\alpha_v[b(x)]\,dxdydudv\cr
&= \int \det[T^*]^{-1}e(T^{-1}u,-v)e(T^{-1}u,y)a(y) e(T^{-1}v,x)b(x)\,dxdydudv\cr
&= \int \det[T^*]^{-1}e(T^{-1}u,y-v)a(y) e(T^{-1}v,x)b(x)\,dxdydudv\cr
\end{align*}
The integral of  $e({T^*}^{-1}u,v-y)$ over $u$ produces a delta function concentrated on $v=-y$, and then the $v$ integral gives
\begin{align*}
(I(a)\star I(b)) &= \int e(T^{-1}y,x)a(y) b(x)\,dudv \\
&= \int e(T^{-1}x,y)a(y) b(x)\,dudv \\
&= I(a*b),
\end{align*}
 showing that $I$ defines a homomorphism from the Kasprzak deformation to the Rieffel deformation.
\end{proof}

 Standard harmonic analysis shows that this is formally an isomorphism on suitably defined smooth subalgebras. (The inverse map takes an algebra element $a$ and does harmonic analysis of $\alpha$ action setting $a(x)$ to be the component of $a$ such that 
 $\alpha_y[a(x)] = \sigma^{-1}(x,y)a(x)$.) The same constructions can be carried out for 
 $C_0(X)$-algebras.

\section{Classifying noncommutative principal torus bundles}
\label{sect:classify}

The noncommutative principal torus bundles of Echterhoff, Nest, and
Oyono-Oyono, whose definition was recalled in Section 1, were classified in
\cite{ENOO} and will be outlined in this section. 
We also give a classification of fibrewise smooth
noncommutative principal torus bundles in terms of parametrized
strict deformation quantization of ordinary principal torus bundles.

By Takai--Takesaki duality $\alg(X)$ is Morita equivalent to
$C_0(X, \cpt)\rtimes \ft{T}$, so the authors in \cite{ENOO} note that the NCPT-bundles can be
classified by up to Morita equivalence by the outer equivalence
classes ${\mathcal E}_{\ft{T}}(X)$ of $\ft{T}$-actions, and one
has the sequence
$$
0 \longrightarrow H^1(X, T) \longrightarrow {\mathcal E}_{\ft{T}}(X)
\longrightarrow  C(X,H^2(\ft{T},\torus)) \longrightarrow 0.
$$
This leads to a classification in terms of  a principal torus
bundle $q:Y \to X$, from $H^1(X,T)$, and a map $\sigma \in 
C_b(X, H^2(\ft{T},\torus))$, the equivalence classes of
multipliers on the dual group  $\ft{T}$.
These data define a noncommutative torus bundle by forming the
fixed point algebra
$$
[C_0(Y)\otimes_{C_0(\ft{Z})} C^*(H_\sigma))]^{T}
$$
with $C^*(H_\sigma)$ being the bundle of  group $C^*$-algebras of the
central extensions of $\ft{T}$ by $\ft{Z} :=H^2(\ft{T},\torus)$
defined by $\sigma(x)$ at $x$, the action of $C_0(\ft{Z})$ on
$C_0(Y)$  coming from the composition $\sigma\circ q: Y \to X \to
\ft{Z}$ and that on $C^*(H_\sigma)$ from the natural action of a
subgroup algebra. The bundle is a classical principal bundle
when $\sigma$ is the constant map to the trivial multiplier $1$ (or
indeed is homotopic to any constant map).

A key observation is that the datum $\sigma$, or more practically
an equivalence class of $\sigma\in C(X, Z^2(\ft{T},\torus))$ can be
identified with the similar map in the parametrised deformation
theory, and that the Landstad--Kasprzak dual deformation theory
conveniently matches the duality in the definition of
NCPT-bundles with the group $V = T$ and the dual algebra $\bdd
= C_0(X,\cpt)$. The analysis in \cite{ENOO} starts with the
case of $X$ a point, where the algebra is shown to be the
twisted C$^*$ group algebra of $\ft{T}$ defined by the
multiplier $\sigma$, or, equivalently, the deformed algebra defined
by $\sigma$. The same construction can be carried out in the case of
general $X$ using our parametrised deformation constructions,
and this can then be twisted using an ordinary principal
$T$-bundle. Now given a fibrewise smooth NCPT-bundle
$\alg^\infty(X)$ the defining deformation $\sigma$  can be removed
by a further deformation  by $\conj{\sigma}$ since then one has a
total deformation $\sigma\conj{\sigma} = 1$, and a constant map $1$ gives
an ordinary principal torus bundle up to $T$-equivariant Morita equivalence over $C_0(X)$. 
In other words one can
recover the principal torus bundle $q:Y \to X$ in this way up to 
$T$-equivariant Morita equivalence over $C_0(X)$ via an 
iterated parametrized strict deformation quantization. To
summarize, we have the following main result, which follows from 
Theorem \ref{thm:parameter},  \S\ref{sect: Landstad}, Example \ref{ex:torusbundle},  
and the observations above.

\begin{theorem}\label{thm:classify}
Given a fibrewise smooth NCPT-bundle $\alg^\infty(X)$, there is
a  defining deformation $\sigma\in C_b(X, Z^2(\ft{T},\torus))$ and a
principal torus bundle $q:Y \to X$ such that
$\alg^\infty(X)$ is $T$-equivariant Morita equivalent over $C_0(X)$, to the parametrised strict deformation
quantization of $C^\infty_{\rm {fibre}}(Y)$ (continuous
functions on $Y$ that are fibrewise smooth) with respect to
$\sigma$, that is,
$$
\alg^\infty(X) \cong C^\infty_{\rm {fibre}}(Y)_\sigma.
$$
Conversely, by Example \ref{ex:torusbundle}, the parametrised strict deformation
quantization of $C^\infty_{\rm {fibre}}(Y)$ is the noncommutative 
principal torus bundle $C^\infty_{\rm {fibre}}(Y)_\sigma$.
\end{theorem}


\section{Fine structure of parametrised strict deformation quantization}
\label{sect:examples}
We have seen in Theorem \ref{thm:classify} that all fibrewise smooth NCPT-bundles are just parametrised
strict deformation quantizations of ordinary principal torus bundles. We will use this to write out
the fine structure of fibrewise smooth NCPT-bundles.

\begin{example}
We begin by recalling the construction by Rieffel \cite{Rieffel1} realizing the smooth noncommutative
torus as a deformation quantization of the smooth functions on a torus
$T = \reals^{n}/\integer^{n}$ of dimension equal to $n$.

Recall that any translation invariant Poisson bracket on $T$  is just 
$$
\{a, b\} = \sum \theta_{ij} \frac{\partial a}{\partial x_i}\frac{\partial b}{\partial x_j},
$$
for $a, b \in C^\infty(T)$, where $(\theta_{ij} )$ is a skew symmetric matrix.
The action of $T$ on itself is given by translation. The Fourier transform is an isomorphism
between $C^\infty(T)$ and $\cS(\hat T)$, taking the pointwise product on $C^\infty(T)$ to the
convolution product on $\cS(\hat T)$ and taking differentiation with respect to a coordinate function
to multiplication by the dual coordinate. In particular, the Fourier transform of the Poisson bracket gives
rise to an operation on $\cS(\hat T)$ denoted the same. For $\phi,\psi \in \cS(\hat T)$, define
$$
\{\psi, \phi\} (p)= -4\pi^2\sum_{p_1+p_2=p} \psi(p_1) \phi (p_2) \gamma(p_1, p_2)
$$
where $\gamma$ is the skew symmetric form on $\hat T$ defined by
$$
 \gamma(p_1, p_2) =  \sum \theta_{ij} \,p_{1,i}\,p_{2,j}.
$$
For $\hbar \in \reals$, define a skew bicharacter $\sigma_\hbar$ on $\hat T$ by
$$
\sigma_\hbar(p_1, p_2) = \exp(-\pi\hbar\gamma(p_1, p_2)).
$$
Using this, define a new associative product $\star_\hbar$ on $\cS(\hat T)$,
$$
(\psi\star_\hbar\phi) (p) = \sum_{p_1+p_2=p} \psi(p_1) \phi (p_2) \sigma_\hbar(p_1, p_2).
$$
This is precisely the smooth noncommutative torus $A^\infty_{\sigma_\hbar}$.

The norm $||\cdot||_\hbar$ is defined to be the operator norm for the action of $\cS(\hat T)$
on $L^2(\hat T)$ given by $\star_\hbar$. Via the Fourier transform, carry this structure
back to $C^\infty(T)$, to obtain the smooth noncommutative torus
as a strict deformation quantization of $C^\infty(T)$, \cite{Rieffel1} with respect to the
translation action of $T$.
\end{example}

\begin{example}\label{ex:torusbundle}
We next generalize the above to the case of principal torus bundles $q:Y\to X$ of rank equal to $n$.
Note that fibrewise smooth functions on $Y$ decompose as a direct sum,
\begin{align*}
C^\infty_{\rm{fibre}}(Y) &= \widehat\bigoplus_{\alpha \in \hat T} \ C^\infty_{\rm{fibre}}(X, \cL_\alpha) \\
\phi &= \sum_{\alpha \in \hat T} \phi_\alpha
\end{align*}
where $ C^\infty_{\rm{fibre}}(X, \cL_\alpha)$ is defined as the subspace of $C^\infty_{\rm{fibre}}(Y) $ 
consisting of functions which transform under the character $\alpha \in \hat T$, and where $ \cL_\alpha$ 
denotes the associated line bundle $Y \times_\alpha \complex$ over $X$. That is, 
$ \phi_\alpha(yt) = \alpha(t)  \phi_\alpha(y), \, \forall\, y\in Y,\, t\in T$.
The direct sum is completed in such a way that the function 
$\hat T \ni \alpha \mapsto ||\phi_\alpha||_\infty \in \reals$ 
is in $\cS(\hat T)$. In this interpretation of $C^\infty_{\rm{fibre}}(Y)$, it is easy to extend to this case, the
explicit deformation quantization given in the previous example, which we now briefly outline.
For $\phi, \psi \in C^\infty_{\rm{fibre}}(Y)$, define a new associative product $\star_\hbar$ on $C^\infty_{\rm{fibre}}(Y)$ as follows. 
For $y \in Y$, $\alpha, \alpha_1, \alpha_2 \in \hat T$, let
$$
(\psi\star_\hbar\phi) (y, \alpha) = \sum_{\alpha_1\alpha_2=\alpha} \psi(y, \alpha_1) 
\phi (y, \alpha_2) 
\sigma_\hbar(q(y); \alpha_1, \alpha_2),
$$
using the notation $ \psi(y, \alpha_1) =  \psi_{\alpha_1}(y)$ etc., and where 
$\sigma_\hbar \in C_b(X,  Z^2(\hat T, \torus))$ is a continuous 
 family of bicharacters of
$\hat T$ such that $\sigma_0 =1$, which is part of the data that we start out with. 
We remark that one way to get such a 
$\sigma_\hbar$ is to choose a  continuous family skew-symmetric forms
on $\hat T$, $\gamma: X \longrightarrow Z^2(\hat T, \reals)$, and define 
$\sigma_\hbar = \exp(-\pi \hbar \gamma)$. In the case of the principal torus bundle $Y$, 
we note that the vertical tangent bundle 
of $Y$ has a Poisson structure, i.e. $\gamma \in \bigwedge^2 T^{vert}Y$, which can
be naturally interpreted as a continuous family of symplectic structures along the fibre, that is, $\gamma$
is of the sort considered just previously.
We denote the deformed algebra 
by  $C^\infty_{\rm{fibre}}(Y)_\hbar$, and using \S \ref{sect:parameter}, we can realize it 
as a parametrised strict deformation quantization of $C^\infty_{\rm{fibre}}(Y)$. Since the 
construction is $T$-equivariant, $C^\infty_{\rm{fibre}}(Y)_\hbar$ has a $T$-action that is 
induced from the given $T$-action on $C^\infty_{\rm{fibre}}(Y)$.
\end{example}

\begin{example}
We next consider the example which was one of the inspirations
for Theorem \ref{thm:classify}. Although it is a special case
of the previous example, and is probably also treated
elsewhere, nevertheless we think that it is worthwhile to treat
in our context. Consider a 3-dimensional torus, which we write
as $S^1 \times T$, where $T$ is a two dimensional torus. Let
$\{\cdot, \cdot\}$ denote the Poisson bracket on $S^1\times T$
coming from $T$ and trivial on $S^1$. Then this Poisson bracket
is invariant under the $T$ action on $S^1\times T$, where $T$
acts trivially on $S^1$ and via translation on itself. Here the
fibres are $T$. As in the example above, we construct a strict
deformation quantization of $C^\infty_{\rm{fibre}}(S^1\times
T)$. Taking the partial Fourier transform in the $T$-variables,
we obtain an isomorphism between
$C^\infty_{\rm{fibre}}(S^1\times T)$ and
$\cS_{\rm{fibre}}(S^1\times \hat T)$. In the notation of the
previous example, for $\phi, \psi \in
\cS_{\rm{fibre}}(S^1\times \hat T)$, define
$$
(\psi \star_\hbar \phi)(y, p) =  \sum_{p_1+p_2=p} \psi(y, p_1) \phi (y, p_2) \sigma_\hbar(y; p_1, p_2).
$$
where $\sigma_\hbar :  S^1=\reals/\integer \longrightarrow H^2(\hat T, \torus)\cong \torus$ is the
family of bicharacters of $\hat T$ given by
$$
\sigma_\hbar(y; p_1, p_2) = \exp(-\pi\hbar y\gamma(p_1, p_2)).
$$
Here $\gamma$ is defined as in Example 6.1.
This gives us a family of smooth noncommutative tori, that is,
$$
 \cS_{\rm{fibre}}(S^1\times \hat T)_\hbar = \int_{y\in S^1} A^\infty_{\sigma_\hbar(y)}
$$
which in turn can be identified with (when $\hbar=1$) the
fibrewise Schwartz subalgebra of the 3-dimensional integer
Heisenberg group, ${\rm Heis}_\integer$. That is, the norm
closure of $ \cS_{\rm{fibre}}(S^1\times \hat T)_{\hbar=1}$ is
isomorphic to $C^*({\rm Heis}_\integer)$.

On the other hand, 
using the results of \S \ref{sect:parameter}, we see that 
$ \cS_{\rm{fibre}}(S^1\times \hat T)_\hbar$ is a parametrised strict
deformation quantization of 
$C^\infty_{\rm{fibre}}(S^1\times T)$.
\end{example}

\begin{example}
Motivated by Theorem  \ref{thm:classify} and an example in
\cite{Rieffel1}, we define noncommutative non-principal torus
bundles as follows. Let $\rho: \pi_1(X)\to {\rm
Sp}(2n, \integer)$ be a representation of the fundamental group,
$T = \reals^{2n}/\integer^{2n}$ be the torus and $q: Y_\rho \to X$ be the
non-principal torus bundle given by $Y_\rho = ({\widetilde X}
\times T)/\pi_1(X)$, where we observe that the symplectic group
is a subgroup of the automorphism group of $T$ and $\Gamma=\pi_1(X)$
acts on $T$ via $\rho$ and on the universal cover $r:{\widetilde X}\longrightarrow X $ via
deck transformations.

Let $\{\cdot, \cdot\}$ denote the Poisson bracket on
${\widetilde X}\times T$ coming from $T$ and trivial on
${\widetilde X}$. Then this Poisson bracket is invariant under
the $T$ action on ${\widetilde X}\times T$, therefore
descending to a Poisson bracket on ${\widetilde X}$, denoted by
the same symbol. As in the example above, we construct a strict
deformation quantization of $C^\infty_{\rm{fibre}}({\widetilde
X}\times T)$, which is the algebra of continuous functions on
${\widetilde X}\times T$ that are smooth along the fibres.
Taking the partial Fourier transform in the $T$-variables, we
obtain an isomorphism between
$C^\infty_{\rm{fibre}}({\widetilde X}\times T)$ and
$\cS_{\rm{fibre}}({\widetilde X}\times \hat T)$. In the
notation of the previous example, for $\phi, \psi \in
\cS_{\rm{fibre}}({\widetilde X}\times \hat T)$, define
$$
(\psi \star_\hbar \phi)(y, p) =  \sum_{p_1+p_2=p} \psi(y, p_1) \phi (y, p_2) \sigma_\hbar(r(y); p_1, p_2).
$$
where $\sigma_\hbar :  X \longrightarrow Z^2(\hat T, \torus)$ is a continuous
family of bicharacters of $\hat T$, which is part of the data that we start out with.

Then transporting this structure back to
$C^\infty_{\rm{fibre}}({\widetilde X}\times T)$ gives a strict
deformation quantization such that $\Gamma=\pi_1(X)$
acts properly on it, cf. \cite{Rieffel1}. We denote the deformed algebra as
$C^\infty_{\rm{fibre}}({\widetilde X}\times T)_\hbar$. 
The fixed point subalgebra $C^\infty_{\rm{fibre}}({\widetilde
X}\times T)_\hbar^\Gamma$ of the deformed algebra is then the desired
parametrised strict deformation quantization,
$C^\infty_{\rm{fibre}}({\widetilde X}\times T)_\hbar^\Gamma =
C^\infty_{\rm{fibre}}(Y_\rho)_{\sigma_\hbar}$, where we note
that $C^\infty_{\rm{fibre}}({\widetilde X}\times T)^\Gamma=
C^\infty_{\rm{fibre}}(Y_\rho)$. This is our
definition of a noncommutative non-principal torus bundle. To
summarize, it is determined by
 two pieces of data:
 \begin{itemize}
\item  $\rho \in {\rm Hom}(\pi_1(X),  {\rm Sp}(2n, \integer))$;
\item $\sigma \in C(X, Z^2(\hat T, \torus))$, that is, a continuous family of bicharacters of $\hat T$.
\end{itemize}
\end{example}

\appendix
\section{Factors of automorphy}

Appendix C to \cite{BEM} introduced a method for lifting
algebra bundles to a contractible universal cover and encoding
information about the Dixmier--Douady class in a factor of
automorphy $\ft{j}$. This also fits into a parametrised
deformation picture, but with the further generalisation that
the group $\Gamma$ now acts on the parameter space $X$. The
cocycle $\ft{j}$ for a lifting can be reconstructed from the
Dixmier--Douady class $\delta\in H^3(X,\integer) \sim
H^3(\Gamma,\integer)$, by first finding $\ft{\tau}(k_1,k_2,x)$
($k_1,k_2 \in \Gamma$, $x\in X$) with $d\ft{\tau} = \delta$,
defining $\tau = \exp(2\pi i\ft{\tau})$, and then finding
$\ft{j}(k,x)$ satisfying
$$
\ft{j}(k_1,k_2x)\ft{j}(k_2,x) = \tau(k_1,k_2,x)\ft{j}(k_1k_2,x),
$$
which can be achieved by a modified $\tau$-inducing construction, which gives $\ft{j}$ in terms of $\tau$.

We know that $\tau$ is a $C_0(X)$-valued cocycle  satisfying the cocycle condition
$$
\tau(k_1k_2,k_3)\alpha_{k_3}^{-1}[\tau(k_1,k_2)] = \tau(k_1,k_2k_3)\tau(k_2,k_3)
$$
where $\alpha$ just gives the translation action on $C_0(X)$, and similarly suppressing the $X$-dependence in $\ft{j}$ allows us to rewrite its cocycle condition as
$$
\alpha_{k_2}^{-1}[\ft{j}(k_1)]\ft{j}(k_2) = \tau(k_1,k_2)\ft{j}(k_1k_2).
$$
The cocycle condition on $\tau$ can also be written as
$$
\alpha_{k_3}[\tau(k_1k_2,k_3)]\tau(k_1,k_2) = \alpha_{k_3}[\tau(k_1,k_2k_3)]\alpha_{k_3}[\tau(k_2,k_3)]
$$
so setting
$U(k_1):  k \mapsto  \alpha_k[\tau(k_1,k)]$
we get
$$
U(k_1k_2)(k_3)\tau(k_1,k_2) = \alpha_{k_2}^{-1}[U(k_1)(k_2k_3)]U(k_2)(k_3).
$$
We can lift the automorphism  $\alpha_{k_2}$ to
$$
\wt{\alpha}_{k_2}^{-1}[U(k_1)](k_3)]  =  \alpha_{k_2}^{-1}[U(k_1)(k_2k_3)]
$$
and then
$$
U(k_1k_2)\tau(k_1,k_2) = \wt{\alpha}_{k_2}^{-1}[U(k_1)]U(k_2),
$$
the type of cocycle condition to be satisfied by $\ft{j}$.

To compare these with the Landstad--Kasprzak construction we take $\Gamma = \ft{V}$,
$\ft{\rho}$ the left translation $(L_kf)(x) = f(k^{-1}x)$. Now we think of $\ft{j}$ as a map from $K$ to unitary multipliers on $C_0(X)$, and take $U(k): x \mapsto \ft{j}(k,k^{-1}x)$,  noting that the cocycle condition $\ft{j}(k_1,k_2x)\ft{j}(k_2,x) = \tau(k_1,k_2,x)\ft{j}(k_1k_2,x)$ gives
$$
U(k_1)\ft{\rho}(k_1)[U(k_2)] = \tau(k_1,k_2)U(k_1k_2),
$$
precisely the condition arising in deformation (though the new ingredient is that $K$ acts on the $X$ argument of $\tau$).

We note also that the predual algebra in the Landstad theory is the generalised fixed point algebra $\alg = \bdd^{\ft{\rho}}$.


\begin{thebibliography}{99}

\bibitem{BEM}
P.  Bouwknegt, K.C.\ Hannabuss, and V. Mathai, 
{\em C$^*$-algebras in tensor categories},
 {Clay Mathematics Proceedings}.
{\bf 12} (2009) 39 pages, (in press).  [math.QA/0702802]

\bibitem{Co80}
A.\ Connes,  
{\em C$^*$-alg\`ebres et g\'eometrie diff\'erentielle},
C.R. Acad. Sci. Paris, Ser. A-B, {\bf 290}, (1980) no. 13, 599--604.


 \bibitem{ENOO}
S.\  Echterhoff,  R.\  Nest, and H. Oyono-Oyono,  
{\em  Principal non-commutative torus bundles}, 
Proc. London Math. Soc. (3) {\bf 99},  (2009) 1--31.

\bibitem{EW}
S.\ Echterhoff and D.\ P.\ Williams,
{\em Crossed products by $C\sb 0(X)$-actions},
J.\ Funct.\ Anal.\ \textbf{158} (1998) no.\ 1, 113--151.

\bibitem{KCH}
K.C.\ Hannabuss, 
{\em Representations of nilpotent locally compact groups}, 
J.\ Funct.\ Anal. {\bf 34} (1979) no. 1, 146--165. 


 \bibitem{HRW}
 A.\  an Huef, I.\  Raeburn, and D.P.\  Williams, 
{\em Functoriality of Rieffel's generalised fixed point algebras for proper actions}, 
 [arXiv:0909.2860].

\bibitem{KQ}
S.\  Kaliszewski and J.\  Quigg, 
{\em Categorical Landstad duality for actions}, 
Indiana Math. J. {\bf 58} (2009) 415--441.

\bibitem{Kas88} G. Kasparov, 
\emph{Equivariant $K$-theory and the Novikov conjecture}, 
Invent. Math.   \textbf{91}, (1988) 147--201.

\bibitem{Kasp}
P.\  Kasprzak,
{\em Rieffel deformation via crossed products}, 
J.\ Funct.\ Anal. {\bf 257} (2009) 1288--1332.

\bibitem{Kl} A.\ Kleppner, 
{\em Multipliers on abelian groups}, 
Math. Ann. {\bf 158} (1965) 11--34.

\bibitem{Land1}
M.B. Landstad,
{\em Duality theory for covariant systems}, 
Trans. Amer. Math. Soc. {\bf 248} (1979) 223--267.

\bibitem{Land2}
M.B.\  Landstad, 
{\em Quantization arising from abelian subgroups}, 
Internat. J. Math. {\bf 5} (1994) 897--936.

\bibitem{MR}
V. Mathai and J. Rosenberg,
{\em T-duality for torus bundles via noncommutative topology},
Commun. Math. Phys.,  {\bf 253}  no. 3 (2005) 705--721. [hep-th/0401168]

\bibitem{MR2}
V. Mathai and J. Rosenberg,
{\em T-duality for torus bundles with H-fluxes via noncommutative topology, II: the high-dimensional case and the T-duality group},
Adv. Theor. Math. Phys.,  {\bf 10} no. 1 (2006) 123--158.  [hep-th/0508084]

\bibitem{Rieffel1}
M.A.\  Rieffel,  
{\em Deformation quantization for actions of ${\bf R}^d\,$}, 
Memoirs of the Amer. Math. Soc.  106  (1993),  no. 506, 93 pp. 

\bibitem{Rieffel2}
M.A.\  Rieffel,  
{\em Quantization and C$^*$-algebras}, 
Contemporary Math. {\bf 167},  (1994), 67--97.

\bibitem{Rieffel3}
M.A.\  Rieffel,  
{\em Applications of strong Morita equivalence to transformation group $C^{\ast} $-algebras}, 
Proceedings of Symposia in Pure Mathematics, {\bf 38} (1982) Part I, 299--310. 

\end{thebibliography}
\end{document}